\documentclass[aps,pra,twocolumn,groupedaddress,showpacs,floatfix]{revtex4}
\usepackage{graphicx}
\usepackage{amsmath}

\begin{document}

\title{Shear viscosity of a universal Fermi gas near the superfluid phase transition}

\author{J. A. Joseph$^1$, E. Elliott$^{1,2}$,  and J. E. Thomas$^1$}

\affiliation{$^{1}$Department of  Physics, North Carolina State University, Raleigh, NC 27695, USA}
\affiliation{$^{2}$Department of Physics, Duke University, Durham, NC 27708, USA}

\date{\today}

\begin{abstract}
We precisely measure the shear viscosity for a resonantly interacting Fermi gas as a function of temperature, from nearly the ground state through the superfluid phase transition at a critical temperature $T_c$.  Using an iterative method to invert the data, we extract  the {\it local} shear viscosity coefficient $\alpha_S(\theta)$ versus reduced temperature $\theta$,  revealing previously hidden features. We find that $\alpha_S$ begins to decrease rapidly with decreasing $\theta$ well above $T_c$, suggesting that  preformed pairs play an important role. Further, we observe that the derivative $\alpha_S'(\theta)$ has a maximum at $T_c$.  We compare the local data to several microscopic theories. Finally, we determine the local ratio of the shear viscosity to the entropy density.
\end{abstract}

\maketitle

Condensates of bosons or fermion pairs exhibit nearly frictionless hydrodynamic flow near and below a critical temperature, $T_c$, which is a defining and striking macroscopic property of superfluids. Just above $T_c$, where the fluid is normal, a regime of extremely small, but finite, shear viscosity is observed. A universal lower bound for the ratio of shear viscosity to entropy density of $\hbar/(4\pi\,k_B)$ is conjectured for this normal fluid regime~\cite{Kovtun}. Below $T_c$, the behavior of the shear viscosity of bosonic and fermionic fluids  is quite different. In bosonic $^4$He,  there is an increase in the shear viscosity as the temperature decreases below $T_c$, which is believed to arise from single particle bosonic excitations that couple to the collective (Nambu-Goldstone) modes~\cite{Levin,LevinViscosity}.  In fermionic $^3$He, the shear viscosity decreases rapidly to zero as the temperature decreases below $T_c$, most likely as a result of the suppression of fermionic excitations at low temperatures.  This is consistent with the BCS theory of weakly interacting Fermi superfluids, where there is no coupling to the Nambu-Goldstone boson modes~\cite{LevinViscosity}.

An optically trapped, ultra-cold Fermi gas of atoms tuned near a collisional (Feshbach) resonance provides a new paradigm for the study of shear viscosity in quantum fluids~\cite{CaoViscosity,CaoNJP}, enabling experimental access not only to Bose and Fermi superfluid systems, but also to a resonant, universal regime, where the gas has both fermionic and bosonic properties. Near a Feshbach resonance~\cite{BartensteinFeshbach,JochimFeshbach}, a bias magnetic field applied to a trapped cloud tunes the interaction strength between atoms in two different hyperfine states, denoted spin-up and spin-down.  Well above resonance, atoms in different spin states are weakly attractive, and the system can be described by Bardeen-Cooper-Schrieffer (BCS) theory.  Well below resonance, pairs of spin-up and spin-down atoms are tightly bound into weakly repulsive molecular bosons, where Bose-Einstein condensate (BEC) theory is applicable.  On resonance, the trapped cloud is a very strongly interacting state of matter, the unitary or universal Fermi gas (UFG).

We report the measurement of the shear viscosity of a UFG as a function of temperature below the superfluid transition temperature, testing the degree to which its transport properties align with those of Bose and Fermi quantum fluids. By observing the expansion of a cigar-shaped cloud, we first obtain the shear viscosity averaged over the density profile.  In this cloud-averaged data, we observe a rapid decrease in the shear viscosity as the temperature is reduced below $T_c$. We then demonstrate a method for inverting the cloud-averaged viscosity data to obtain the {\it local} shear viscosity as a function of reduced temperature, revealing features that were previously hidden in the cloud-averages.  This inverted data for the local shear viscosity is compared to recent theories of the shear viscosity for a UFG in the transition region~\cite{LevinViscosity,BruunViscous,BruunViscousNormalDamping,SchaferViscosity, TaylorViscosity,ZwergerViscosity,DrutShearViscosity,BulgacTempShearUnitary}, which differ in the predicted contributions of pair correlations, fermionic excitations, and bosonic excitations at low temperature. Using the measured local shear viscosity and the measured local entropy density~\cite{KuThermo}, we also determine the local ratio of the shear viscosity to the entropy density, which is compared to the universal lower bound conjectured by Kovtun, Son, and Starinets~\cite{Kovtun}.

In the experiments, a Fermi gas of $^6$Li atoms is prepared in a 50-50 mixture of the two lowest hyperfine states and confined in a cigar-shaped optical trap with an elliptical transverse profile. The trap oscillation frequencies  are $(\omega_x,\omega_y,\omega_z) = 2\pi\times[2210(4),830(2),64(0.5)]$ Hz. The cloud is tuned near a broad Feshbach resonance and cooled by evaporation~\cite{OHaraScience} to nearly the ground state.   The final temperature of the gas is controlled by altering the optical trap lowering curve used for evaporation.

\begin{figure}
\begin{center}\
\includegraphics[width=3.4in]{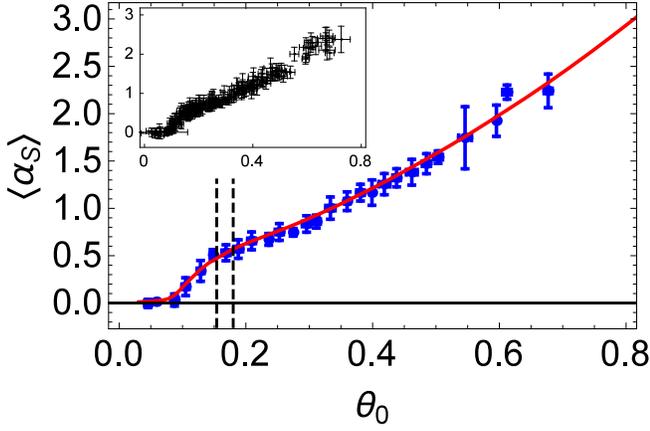}
\end{center}
\caption{Trap-averaged shear viscosity coefficient $\langle\alpha_S\rangle$, where the shear viscosity is $\eta=\alpha_S\,\hbar n$. The solid blue points show the trap-averaged data versus reduced temperature $\theta_0$ at the trap center, after binning in $\theta_0$. The vertical dashed lines denote the critical temperature at the trap center with uncertainty $\theta_c =0.167(13)$~\cite{KuThermo}. The red solid line is obtained by integrating (see Eq.~\ref{eq:trapavcoeff}) the local shear viscosity, which is obtained from the measurements by inverting the raw data (inset).  \label{fig:AvalphaSvsTheta0}}
\end{figure}

The cloud is released from the trap and imaged from two orthogonal directions to determine all three cloud radii $\sigma_i(t)$ at a time $t$ after release. The $\sigma_i(t)$ expand according to $\sigma_i(t)=\sigma_i(0)\,b_i(t)$, where the $b_i(t)$ are hydrodynamic expansion factors that obey universal evolution equations. The hydrodynamic equations depend on the known trap parameters and use the cloud-averaged shear viscosity coefficient $\langle\alpha_S\rangle$ as a free parameter~\cite{ElliottShearViscosity,ElliottScaleInv}. The initial cloud radii $\sigma_i(0)$ and $\langle\alpha_S\rangle$ are self-consistently determined from the transverse aspect ratio $\sigma_x(t)/\sigma_y(t)$ using only {\it one} expansion time $t$ for one measurement~\cite{ElliottShearViscosity}, greatly increasing the data set and energy resolution, see Fig.~\ref{fig:AvalphaSvsTheta0} (inset).

The measured shear viscosity coefficient is related to the shear viscosity, $\eta$, which has a dimension of momentum/area, and hence is given in natural units of $\hbar \,n$, where $n=n({\mathbf{r}})$ is the local density. A dimensionless shear viscosity coefficient $\alpha_S$ is then defined by $\eta\equiv\alpha_S\,\hbar\,n$~\cite{CaoViscosity}. As noted above, the measurements determine a cloud-averaged shear viscosity coefficient $\langle\alpha_S\rangle$, which is defined by
\begin{equation}
\langle\alpha_S\rangle\equiv\frac{1}{N\hbar}\int d^3{\mathbf{r}}\,\eta=\frac{1}{N}\int d^3{\mathbf{r}}\, n\,\alpha_S(\theta),
\label{eq:trapavcoeff}
\end{equation}
where $N$ is the total number of atoms. As shown previously for a UFG, $\alpha_S$ is a function only of the local reduced temperature $\theta\equiv T/T_F(n)$, where $T_F(n)$ is the local Fermi temperature. Further, $\langle\alpha_S\rangle$ is temporally constant as the cloud expands, i.e., it is equal to the {\it trap-averaged} initial value  with $n\rightarrow n({\mathbf{r}},t=0)$~\cite{CaoViscosity,CaoNJP,ElliottShearViscosity}.

Fig.~\ref{fig:AvalphaSvsTheta0} shows the trap-averaged shear viscosity coefficient $\langle\alpha_S\rangle$ as a function of the reduced temperature at the center of the trap $\theta_0 = T/T_F(n_0)$, where $n_0\equiv n({\mathbf{r}=0})$. Temperature is determined from the measured mean square cloud size. Using our known trap potential~\cite{scaled-r} and the equation of state measured by Ku et al.,~\cite{KuThermo}, we determine the local density as a function of reduced temperature at the cloud center, $\theta_0$, i.e., $n(r,\theta_0)$. This relates the measured mean square cloud size to $\theta_0$. As $\theta_0$ is lowered, $\langle\alpha_S\rangle$  decreases rapidly with decreasing temperature below $T_c$.

We now show that the data of Fig.~\ref{fig:AvalphaSvsTheta0} for $\langle\alpha_S\rangle$ versus $\theta_0$ can be inverted to determine the {\it local} shear viscosity coefficient $\alpha_S(\theta)$ as a function of the local reduced temperature $\theta=T/T_F(n)$. In Eq.~\ref{eq:trapavcoeff}, we have $\theta=\theta_0\,(n_0/n)^{2/3}$, so that $\theta$ increases from the minimum $\theta_0$ at the trap center as the density $n$ decreases. This suggests that local shear viscosity can be found from trap-averaged data for different $\theta_0$ by defining  a piecewise representation  $\alpha_i\equiv\alpha_S(\theta_i)$, $i=1...i_{max}$. The $\alpha_i$ are determined  using an iterative matrix inversion and denoising method, which is well-known in imaging processing~\cite{ImageAlgorithm}, and described below.

However, for large $\theta$, $\alpha_S(\theta)\rightarrow \alpha_{3/2}\,\theta^{3/2}\propto T^{3/2}/n$, where $\alpha_{3/2}=45\pi^{3/2}/(64\sqrt{2})\simeq 2.77$~\cite{BruunViscousNormalDamping}. Then, in the low density region, the integrand $n\,\alpha_S(\theta)=n_0\,\alpha_S(\theta_0)$ is independent of density, and the integral is formally divergent. Fortunately, energy conservation assures that the integral must be finite and kinetic theory shows that the shear viscosity $\eta=\hbar\,n\,\alpha_S\rightarrow 0$ as the density vanishes~\cite{BruunViscous}.

We include this behavior  by assuming that the shear viscosity coefficient vanishes abruptly at some effective cut-off radius, $R_c$. We experimentally determine $R_c$ from $\langle\alpha_S\rangle$ data in the temperature region where $\langle \alpha_S \rangle$ has a universal $\theta_0^{3/2}$ dependence~\cite{CaoViscosity,CaoNJP,ElliottShearViscosity}.
Fitting this data using $\langle\alpha_S\rangle=c_0+c_1\,\theta_0^{3/2}$, yields $c_0=0.34(4)$ and $c_1=3.60(15)$~\cite{SupportOnline}. The cutoff radius $R_c$ is then found from Eq.~\ref{eq:trapavcoeff}, which requires $\alpha_{3/2}\,4\pi R_c^3n_0/(3N)=c_1$~\cite{SupportOnline,scaled-r}. We assume a gaussian density profile for the high temperature data, where $n_0 = N(\pi \frac{2}{3} \langle r^2 \rangle)^{-3/2}$, with $\langle{\mathbf{r}}^2\rangle$ the (temperature-dependent) mean square radius of the trapped cloud. Then we find $R_c =0.98\,\langle{\mathbf{r}}^2\rangle^{1/2}$~\cite{SupportOnline}. Making the simplest scale-invariant assumption, we take $R_c =\langle{\mathbf{r}}^2\rangle^{1/2}$ at all temperatures.

Now we assume a piecewise representation of the local shear viscosity, using a discrete chosen set of reduced temperatures $\theta_i$, with  $\alpha(\theta)=\alpha_i$ for $\theta_{i}\leq\theta\leq\theta_{i+1}$. Eq.~\ref{eq:trapavcoeff} is then converted into a system of linear equations, with the $j^{th}$ equation corresponding to the $j^{th}$ measurement of the trap-averaged shear viscosity $\langle\alpha_S\rangle_j$ with a reduced temperature $\theta_{0j}$ at the trap center,
\begin{eqnarray}
\langle \alpha_S \rangle_{j}&=&\sum_i C_{ji} \alpha_i\nonumber\\
C_{ji}&\equiv&\int_{R_{i}(\theta_{0j})} ^{R_{i+1}(\theta_{0j})} 4\pi r^2\,n(r,\theta_{0j})\,dr,
\label{eq:alphamatrix1}
\end{eqnarray}
For each $\theta_i$,  \mbox{$(\theta_i/\theta_{0j})^{3/2}= n(0,\theta_{0j})/n(R_i,\theta_{0j})$} determines $R_i(\theta_{0j})$.
We can write Eq.~\ref{eq:alphamatrix1} in matrix form,
\begin{equation}
\mathbf{\langle \boldsymbol\alpha \rangle} = \mathbf{C} \cdot \mathbf{\boldsymbol\alpha} +\boldsymbol\Delta\alpha,
\label{eq:alphamatrix2}
\end{equation}
where we have added a  vector $\boldsymbol\Delta\alpha$ to represent the noise in the fit arising from imperfect data.

Borrowing from image analysis procedures~\cite{ImageAlgorithm}, we use an iterative method coupled with denoising techniques to solve Eq.~\ref{eq:alphamatrix2}.  We find that this method provides the best removal of high frequency noise associated with measurements, but leaves enough resolution to determine  the smooth behavior and significant transitions in the local shear viscosity.  The iterative solution takes the general form~\cite{ImageAlgorithm},
\begin{equation}
\mathbf{\boldsymbol\alpha_{m+1}} = (1-\beta)\mathbf{\boldsymbol\alpha_m} + \beta \Psi[\mathbf{\boldsymbol\alpha_m}+\mathbf{C^T}(\mathbf{\langle \boldsymbol\alpha \rangle}-\mathbf{C} \cdot \mathbf{\boldsymbol\alpha_m})]
\end{equation}
where $m$ is the iteration number and $\boldsymbol\alpha_{m+1}$ is determined from the previous $m$-step, $\boldsymbol\alpha_{m}$. Here, $0 \le \beta \le 1$ is an adjustable parameter that determines the speed of convergence of the iterative process, $\mathbf{C^T}$ is the transpose of $\mathbf{C}$, and $\Psi({\mathbf{x}})$ is a denoising or smoothing function.  For our purpose, we have found that choosing $\Psi({\mathbf{x}})$ to be a simple three-point moving average provides sufficient denoising.  Further, we require that $\alpha_i$ monotonically increase as a function of reduced temperature $\theta_i$, as suggested in Ref.~\cite{LevinViscosity}.  The local shear viscosity converges slowly from an initial seed $\boldsymbol\alpha$, which we take to be the high temperature approximation for the local shear viscosity, discussed above, $\alpha_S (\theta) = 2.77\,\theta^{3/2}$.

For our inversion, we find that this method is robust in the choice of $\beta$~\cite{SupportOnline}.   We monitor the change in $\boldsymbol\alpha$ as a function of iteration number $m$.  When the change is sufficiently small we stop the algorithm~\cite{SupportOnline}.  For the data presented in this paper we have chosen  $\beta=0.1$.  The algorithm converges after only 12 iterations. The supplemental material provides a review of the inversion  methods~\cite{SupportOnline}. As a consistency check, we integrate the local shear viscosity obtained from the algorithm over the cloud volume using Eq.~\ref{eq:trapavcoeff}, yielding the red curve shown in Fig.~\ref{fig:AvalphaSvsTheta0},  which agrees very well with the measured trap-averaged viscosity coefficients.

\begin{figure}
\begin{center}\
\includegraphics[width=3.25in]{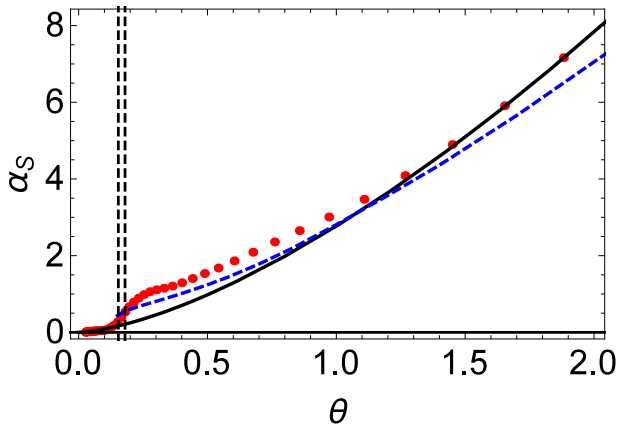}
\end{center}
\caption{Local shear viscosity coefficient $\alpha_S(\theta)$ versus reduced temperature $\theta=T/T_F(n)$, where the local shear viscosity is $\eta=\alpha_S\,(\theta)\hbar n$.  Red dots show the experimental results obtained from our data inversion method. The vertical dashed lines denote the critical reduced temperature with uncertainty $\theta_c=0.167(13)$~\cite{KuThermo}. Black-solid line, Boltzmann limit, $\alpha = 2.77 \,\theta^{3/2}$ from Bruun and Smith~\cite{BruunViscousNormalDamping}; Blue-dashed curve, prediction by Enss et al., Ref.~\cite{ZwergerViscosity}.  \label{fig:alphaSvsThetaHigh}}
\end{figure}

Fig.~\ref{fig:alphaSvsThetaHigh} shows the local shear viscosity as a function of reduced temperature $\theta$ emphasizing the higher temperature regime. At the highest temperatures shown, the local shear viscosity is consistent with the two-body Boltzmann equation limit $\alpha = 2.77\,\theta^{3/2}$~\cite{BruunViscousNormalDamping}, which is shown as the solid black curve that falls below $\alpha_S$ as the temperature decreases.  The blue-dashed curve shows the prediction of Enss, Haussman, and Zwerger~\cite{ZwergerViscosity}, which approximately captures the curvature for $\theta <1.5$, but is below the Boltzmann limit at and above $\theta =2$.

\begin{figure}
\begin{center}\
\includegraphics[width=3.25in]{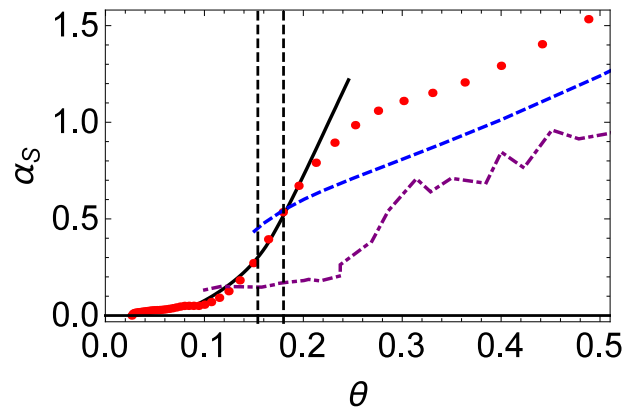}
\end{center}
\caption{Local shear viscosity coefficient $\alpha_S(\theta)$ versus reduced temperature $\theta=T/T_F(n)$ near the superfluid transition region, where the local shear viscosity is $\eta=\alpha_S\,(\theta)\hbar n$.  Red dots show the experimental results obtained from our data inversion method. The vertical dashed lines denote the critical temperature with uncertainty $\theta_c=0.167(13)$~\cite{KuThermo}. Blue-dashed curve from Enss et al,  Ref.~\cite{ZwergerViscosity}. Black-solid line from Guo et al., Ref.~\cite{LevinViscosity}, showing very good agreement. Purple dot-dashed curve from Wlazlowski et al., Ref.~\cite{DrutShearViscosity}. \label{fig:alphaSvsThetaLow}}
\end{figure}
The local shear viscosity reveals important features that are hidden in the trap-averaged data.  Fig.~\ref{fig:alphaSvsThetaLow} shows $\alpha_S$ in the low temperature regime. The rapid decrease in $\alpha_S$ with decreasing temperature begins well above $T_c$, suggesting that preformed pairs are important. Further, in the region $T<T_c$, our measured of local shear viscosity is in remarkably good agreement with theoretical predictions based generally on a pseudogap-BCS theory, which includes such non-condensed pairs~\cite{LevinViscosity}.  The QMC~\cite{DrutShearViscosity} results capture the general shape but not the absolute scale of the data. At the very lowest temperatures measured, $\alpha_S$ is consistent with zero.

 \begin{figure}
\begin{center}\
\includegraphics[width=3in]{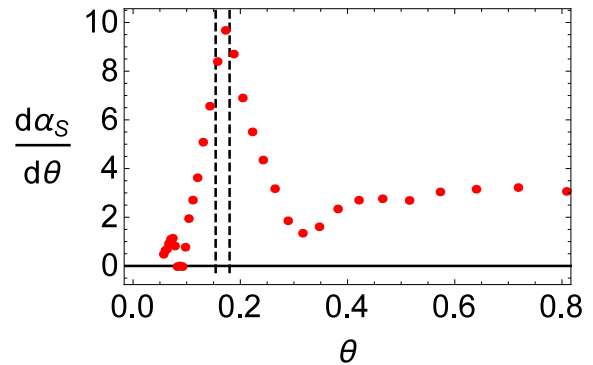}
\end{center}
\caption{Slope of the shear viscosity coefficient $\alpha_S$ versus reduced temperature $\theta$, showing a peak at $T_c$. The vertical dashed lines denote the critical reduced temperature with uncertainty $\theta_0=0.167(13)$~\cite{KuThermo}.\label{fig:alphaslope}}
 \end{figure}
 
We find the interesting result that the slope $d \alpha_S / d \theta$ of the inverted data has a peak at the superfluid transition temperature, Fig.~\ref{fig:alphaslope}, which is robust with respect to our choice of parameters in implementing the data inversion. This directly reveals the local superfluid transition, as observed previously in the heat capacity~\cite{KuThermo} and by Bragg scattering~\cite{ValeLocalBraggScatt}.
\begin{figure}
\begin{center}\
\includegraphics[width=3in]{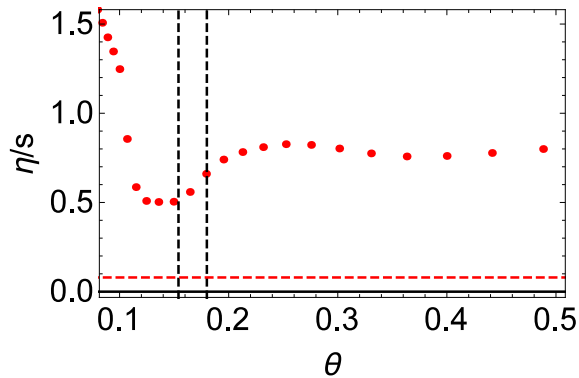}
\end{center}
\caption{Ratio of shear viscosity $\eta$ to the entropy density $s$, in units of $\hbar/k_B$, versus reduced temperature $\theta=T/T_F(n)$. Red dots are the ratio of the local shear viscosity, obtained from our matrix inversion method, to the entropy density measured in Ref.~\cite{KuThermo}.  The vertical dashed lines denote the critical reduced temperature with uncertainty $\theta_0=0.167(13)$~\cite{KuThermo} and the horizontal dashed line at $1/(4\pi)$ indicates the KSS lower bound~\cite{Kovtun}. \label{fig:EtaOverSvsTheta}}
\end{figure}
Next, we determine the ratio of the local shear viscosity to the local entropy density, using the entropy data of Ref.~\cite{KuThermo}. The ratio is compared to the lower bound conjectured by Kovtun, Son, and Starinets~\cite{Kovtun}, as shown in Fig.~\ref{fig:EtaOverSvsTheta}.  We find that for a range of temperature above the superfluid transition temperature, the ratio remains nearly constant.  There appears to be a minimum in the ratio below $T_c$  and an upturn in the ratio as $T\rightarrow 0$.  However, as both the entropy and the viscosity are rapidly approaching zero for $\theta<0.1$, the error associated with both of the measured quantities does not permit an unambiguous determination of the behavior in this very low temperature regime.

The observed rapid decrease in the measured shear viscosity below $T_c$  suggests that the universal shear viscosity of a unitary Fermi gas is closer in character to that of fermionic $^3$He than to bosonic $^4$He. The determination of the ratio of the shear viscosity to the entropy density at the lowest temperatures will require improved precision in the measurement of both quantities.

This research is supported by the Physics Division of the
 National Science Foundation (Quantum hydrodynamics in interacting Fermi gases) and by the
 Division of Materials Science and Engineering,  the
Office of Basic Energy Sciences, Office of Science, U.S.
Department of Energy (Thermodynamics in strongly correlated Fermi gases). Additional support
has been provided by the Physics Divisions of the Army Research Office and the Air Force Office of Scientific Research.  The authors are pleased to acknowledge M. Bluhm and T. Sch\"{a}fer, North Carolina State University, for stimulating conversations and M. Gehm, Duke University, for suggesting the use of image processing methods.


\begin{thebibliography}{22}
\bibitem[{sca()}]{scaled-r}
\bibinfo{note}{We transform to spherical symmetry in the usual way with scaled
  coordinates, $\bar{\omega}\tilde{x}_i\equiv\omega_i x_i$, where
  $\bar{\omega}\equiv(\omega_x\omega_y\omega_z)^{1/3}$, so that $r^2=\sum_i
  \tilde{x}_i^2$, i.e., the effective trap potential energy is then a function
  of $r$. For a harmonic trap, $\sum_i
  m\omega_i^2\,x_i^2/2=m\bar{\omega}^2\,r^2/2$.}
\end{thebibliography}

\widetext
\appendix
\section{Supplemental Material}
\label{sec:supplement}

In this supplemental material, we provide a detailed discussion of the analysis techniques used to determine the local shear viscosity coefficient $\alpha_S$ from the measurements of the trap-averaged shear viscosity coefficient $\langle \alpha_S \rangle$.  First we describe our experimental determination of the cut-off radius $R_c$ from measurements of $\langle \alpha_S \rangle$ at  high temperature.  Then we discuss the iterative procedure we use to solve the inverse linear problem of obtaining $\alpha_S$ from $\langle \alpha_S \rangle$.

\subsection{Cut-Off Radius $R_c$}
\label{Sec:R_c}
We determine $R_c$ experimentally.  As explained in the paper, the the trap-averaged shear viscosity coefficient is given by,
\begin{equation}
\label{eq:S1}
\langle \alpha_S \rangle = \frac{1}{N}\int d^3\mathbf{r} \,n\, \alpha_S(\theta),
\end{equation}
where $n$ is the density and $\theta = T/T_F(n)$ is the reduced temperature, with $T_F(n)$ the local Fermi temperature. At high temperature, where $n\alpha_S(\theta)\propto T^{3/2}$ is density independent, this integral formally diverges.  As discussed in the main paper, the integral must be finite due to energy conservation, and kinetic theory demonstrates that the shear viscosity vanishes as $n \rightarrow 0$.  We include this behavior by assuming that the shear viscosity coefficient vanishes abruptly at some cut-off radius, $R_c$.

At high temperature, we can write $\alpha_S = \alpha_{3/2} \, \theta^{3/2}$.  If we assume that $\alpha_S = 0$ for $r>R_c$, then Eq.~\ref{eq:S1} can be written
\begin{equation}
\label{eq:S2}
\langle \alpha_S \rangle = \alpha_{3/2}\,\theta_0^{3/2} \frac{1}{N}\int_0^{R_c} d^3\mathbf{r}\, n \left(\frac{\theta}{\theta_0}\right)^{3/2},
\end{equation}
where $\theta_0 = T/T_F(n_0)$ is the reduced temperature at the trap center. Since $T_F \propto n^{2/3}$,  $n(\theta/\theta_0)^{3/2} = n_0$. The integral in Eq.~\ref{eq:S2} is then $n_0\,V_c$, where $V_c = \frac{4\pi}{3}R_c^3$, so that
\begin{equation}
\label{eq:S3}
\langle \alpha_S \rangle = \alpha_{3/2}\,\theta_0^{3/2} n_0 \frac{4 \pi R_c^3}{3 N}.
\end{equation}

In Eq.~\ref{eq:S3}, $\langle \alpha_S \rangle$ is known from experimental data~\cite{CaoNJP} and $\alpha_{3/2}=45\pi^{3/2}/(64\sqrt{2})=2.77$~\cite{BruunViscousNormalDamping}.
\begin{figure}[htb]
\begin{center}\
\includegraphics[width=4in]{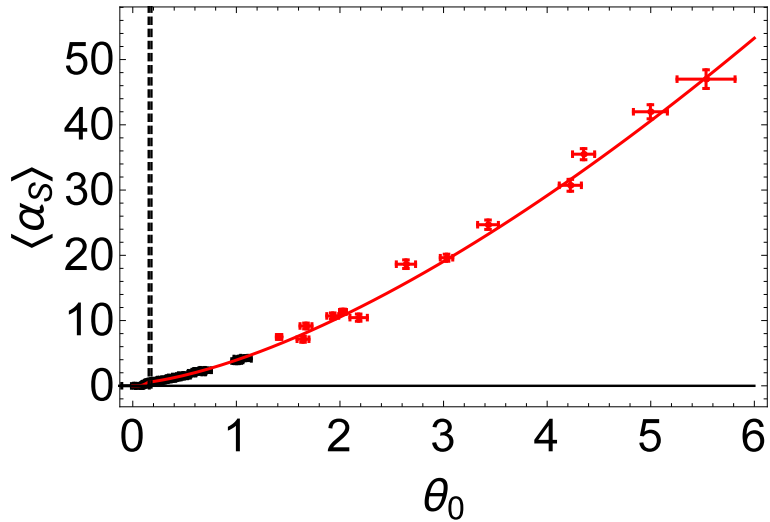}
\end{center}
\caption{Trap-averaged shear viscosity coefficient $\langle \alpha_S \rangle$ versus reduced temperature at the trap center $\theta_0=T/T_F(n_0)$.  Black dots: measurements of the present paper; Red dots: high temperature data of Ref.~\cite{CaoNJP}; Red line: best fit to the data using $c_0 + c_1 \theta_0^{3/2}$. We find $c_0 = 0.34(4)$ and $c_1=3.60(15)$.\label{fig:HighTShear}}
\end{figure}
Fig.~\ref{fig:HighTShear} shows measured $\langle \alpha_S \rangle$ as a function of $\theta_0$, and a fit to the data using $\langle \alpha_S \rangle = c_0 + c_1\, \theta_0^{3/2}$, where $c_0 = 0.34(4)$ and $c_1=3.60(15)$.  For the high temperature limit, we can ignore $c_0$ and solve for $R_c$ using Eq.~\ref{eq:S3}.

We choose to express the cut-off radius as a function of the mean square cloud size $\langle r^2 \rangle$, which increases with the energy of the trapped gas.  The spatial density profile of the gas at high temperature is well fit by a gaussian. Then, the density at the center of the trap is related to the mean square cloud size and atom number by $n_0 = N(\pi \frac{2}{3} \langle r^2 \rangle)^{-3/2}$.  Finally, we obtain the cut-off radius as a function of $\langle r^2 \rangle$,
\begin{equation}
R_c = \left(\sqrt{\frac{\pi}{6}}\frac{c_1}{\alpha_{3/2}}\right)^{1/3}\hspace{-0.1in}\langle r^2 \rangle^{1/2} =\, 0.98 \, \langle r^2 \rangle^{1/2}.
\end{equation}
This result suggests that $R_c = \langle r^2 \rangle^{1/2}$. Note that the cut-off radius is a function temperature and generally depends on the number of atoms and the trap parameters. Remarkably, the cut of radius as determined by a fit to high temperature data is almost exactly the rms cloud size. For inverting the trap-averaged data, we make the simplest scale-invariant assumption and take $R_c = \langle r^2 \rangle^{1/2}$ at all temperatures.

\subsection{Iterative Method}
\label{sec:Iteration}
Iterative matrix inversions are commonly used to solve image restoration and other linear inverse problems~\cite{ImageAlgorithm}.  In this section, we focus on a class of matrix inversion methods that combines a linear problem with a set of non-quadratic regularizers or denoising functions.  First, we clearly state the linear problem and matrix inversion method. Then we discuss the iterative procedure and how we determine when procedure has converged. Finally we explore the effect of our  denoising functions.

\subsubsection{Linear Inverse Problem}
To obtain the local shear viscosity coefficient $\alpha_S$ from the trap-averaged shear viscosity coefficient $\langle \alpha_S \rangle$, the linear inverse problem is relatively simple to state.  As explained in the main paper we use a set of $J=196$ measured shear viscosity coefficients $\langle \alpha_S \rangle_j$ to solve for a set of $I=59$ local shear viscosity coefficients $\alpha_i$.
\begin{equation}
\mathbf{\langle \boldsymbol\alpha \rangle} = \mathbf{C} \cdot \mathbf{\boldsymbol\alpha} +\boldsymbol{\Delta\alpha},
\label{eq:alphamatrix2}
\end{equation}
where $\boldsymbol{\Delta\alpha} = \mathbf{\langle \boldsymbol\alpha \rangle}-\mathbf{C} \cdot \mathbf{\boldsymbol\alpha}$ is the difference between the local result and our measured trap-averaged result due to imperfect data.  The matrix $\mathbf{C}$ is a set of coefficients from the linear equations relating $\alpha_i$ to $\langle \alpha\rangle_j$,
\begin{equation}
\langle \alpha_S \rangle_{j}=\sum_i C_{ji} \alpha_i
\end{equation}
For a square coefficient matrix (i.e. $J = I$) and $\boldsymbol{\Delta\alpha} = 0$ the matrix $\mathbf{C}$ can be inverted to solve for $\boldsymbol\alpha$ in Eq.~\ref{eq:alphamatrix2}.  This is not the case for our experiment.

\begin{figure}[htb]
\begin{center}\
\includegraphics[width=5in]{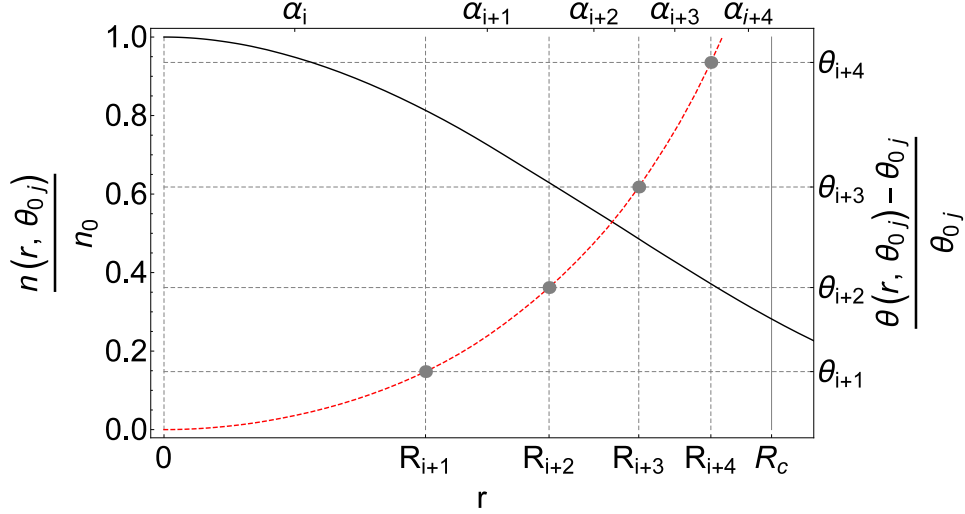}
\end{center}
\caption{Normalized density (black curve) and local reduced temperature (red dashed curve) as a function of radial position.  Each local viscosity coefficient $\alpha_i$ spans the local temperature region $\theta_i$ to $\theta_{i+1}$ and corresponding radii $R_i$ to $R_{i+1}$. \label{fig:VolumeSlices}}
\end{figure}

In order to determine each element in $\mathbf{C}$ we first define the local shear viscosity coefficient vector $\boldsymbol\alpha$ as a function of a local reduced temperature vector $\boldsymbol\theta$.
\begin{equation}
\alpha(\theta) \equiv \left\{
  \begin{array}{ll}
    \alpha_1 &  \theta_1 < \theta \le \theta_2 \\
    \alpha_2 &  \theta_2 < \theta \le \theta_3 \\
    ...\\
    \alpha_I &  \theta_I < \theta \le \theta_{I+1} \\
  \end{array}
\right.
\label{eq:alphaarray}
\end{equation}
where $\boldsymbol\theta$ consists of $I+1$ points and denotes the $\theta_i$ boundaries of $\alpha_i$.  Then for each measurement of $\langle \alpha_S \rangle_j$ and $\theta_{0j}$, we re-construct the density $n(r,\theta_{0j})$ as a function of position $r$ from the measured equation of state~\cite{KuThermo}.  As shown in Fig.~\ref{fig:VolumeSlices} the local reduced temperature for each measurement $\theta(r,\theta_0)$ increases with $r$, and we can identify radii $R_i(\theta_{0j})$ that correspond to a subset of $\boldsymbol\theta$.
Then, the coefficient $C_{ji}$ corresponds to the $j^{th}$ trap-averaged measurement and the $i^{th}$ local coefficient.  $C_{ji}$ is the volume integral of the density $n(r,\theta_{0j})$ between radii $R_i$ to $R_{i+1}$,
\begin{equation}
C_{ji}\equiv\int_{R_{i}(\theta_{0j})} ^{R_{i+1}(\theta_{0j})} 4\pi r^2\,n(r,\theta_{0j})\,dr.
\label{eq:alphamatrix1}
\end{equation}

\begin{equation}
\langle \alpha \rangle_j = \sum_i C_{ji} \alpha_i.
\end{equation}
For this paper we have constructed the list of $\theta_i$, such that there are approximately $10$ equal volume elements for each measurement.  For clarity, only five regions are shown in Fig.~\ref{fig:VolumeSlices}.  Further, for the limits of integration of Eq.~\ref{eq:alphamatrix1}, for the centermost region integral starts at $r=0$ and for the outermost region the integral ends at $R_c$.

\subsubsection{Iterative Matrix Inversion}
In order to solve for the local shear viscosity coefficients $\boldsymbol\alpha$ in Eq.~\ref{eq:alphamatrix2} we borrow from a technique that is commonly used in image processing.  We invert the problem and iteratively solve for the local shear viscosity using the following equation,
\begin{equation}
\mathbf{\boldsymbol\alpha_{m+1}} = (1-\beta)\mathbf{\boldsymbol\alpha_m} + \beta \Psi[\mathbf{\boldsymbol\alpha_m}+\mathbf{C^T}(\mathbf{\langle \boldsymbol\alpha \rangle}-\mathbf{C} \cdot \mathbf{\boldsymbol\alpha_m})],
\label{eq:iteration}
\end{equation}
where $m$ is the iteration number.  $\beta$ determines the speed of convergence and $\Psi$ is a non-quadratic regularizer or denoising function, both of which are discussed in more detail below.  This procedure requires a seed function for the initial value of the local shear viscosity coefficient, $\boldsymbol\alpha_{0} = 2.77 \boldsymbol \theta^{3/2}$.

\subsubsection{Convergence}
Once we have the iterative procedure in place, we need to determine the optimal speed of convergence $\beta$ and a condition of convergence, i.e. at what iteration step $m$ should we stop the algorithm.  We find that the algorithm is stable for all $\beta \lesssim 0.5$, and the result $\boldsymbol\alpha$ is independent of $\beta$.  The only effect of $\beta$ is on the speed of convergence. If $\beta$ is smaller the algorithm takes longer to converge, but the result remains unaffected.  Therefore, we shall continue our discussion of convergence using $\beta = 0.1$.

\begin{figure}[htb]
\begin{center}\
\includegraphics[width=6in]{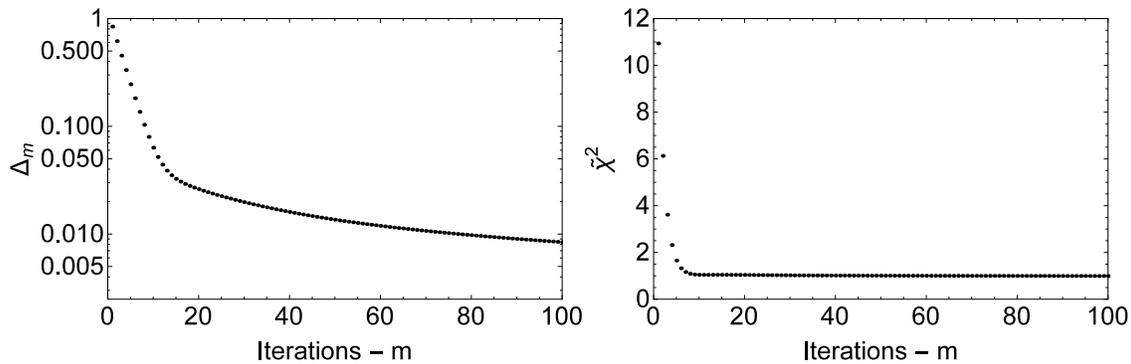}
\end{center}
\caption{The average change in local shear viscosity $\Delta_m$ (left) and goodness of fit $\widetilde{\chi}^2$ (right). We see a distinct change in the slope of $\Delta_m$ at an iteration number $m = 12$  at which point $\widetilde{\chi}^2\simeq 1$. \label{fig:Convergence}}
\end{figure}

To determine convergence, we monitor two parameters during iteration.  One parameter is the rms change in the local shear viscosity for each iteration,
\begin{equation}
\Delta_m  \equiv \frac{1}{\beta}\sqrt{\frac{1}{I}\sum_i(\boldsymbol\alpha_{m} - \boldsymbol\alpha_{m-1})^2}.
\end{equation}
Here we have scaled the rms change in the local shear viscosity by the speed of convergence, $\beta$.  As the local shear viscosity approaches convergence, $\Delta_m$ rapidly decreases.  The other parameter we track is the normalized $\widetilde{\chi}^2$, which determines the goodness of fit of the integrated $\alpha_i$ to our experimental data,
\begin{equation}
\widetilde{\chi}^2 = \frac{1}{J-I}\sum_j \frac{(\mathbf{C} \cdot \boldsymbol\alpha-\langle \alpha_S \rangle_j)^2}{\sigma_{\alpha \,j}^2},
\end{equation}
where $\sigma_\alpha$ is the error in the measured $\langle \alpha_S \rangle$.

Fig.~\ref{fig:Convergence} show $\Delta_m$ and $\widetilde{\chi}^2$ as a function of iteration number.  There is a clear change in the slope of $\Delta_m$ after $12$ iterations, at which point $\widetilde{\chi}^2\simeq 1$.  This means that after $m=12$ iterations of the algorithm, there is very little new information gained by continuing the iterative procedure.  The curves presented in our paper show the iterative matrix solution with $m=12$ iterations.

\subsubsection{Denoising Function $\Psi$}

As stated before, $\Psi({\mathbf{x}})$ is a  denoising function.  We implement $\Psi$ after each iteration by applying a $3$-point moving average to the quantity $\mathbf{\boldsymbol\alpha_m}+\mathbf{C^T}(\mathbf{\langle \boldsymbol\alpha \rangle}-\mathbf{C} \cdot \mathbf{\boldsymbol\alpha_m})$ and by requiring that $\boldsymbol\alpha_m$ increase monotonically.
\begin{figure}[htb]
\begin{center}\
\includegraphics[width=7.0in]{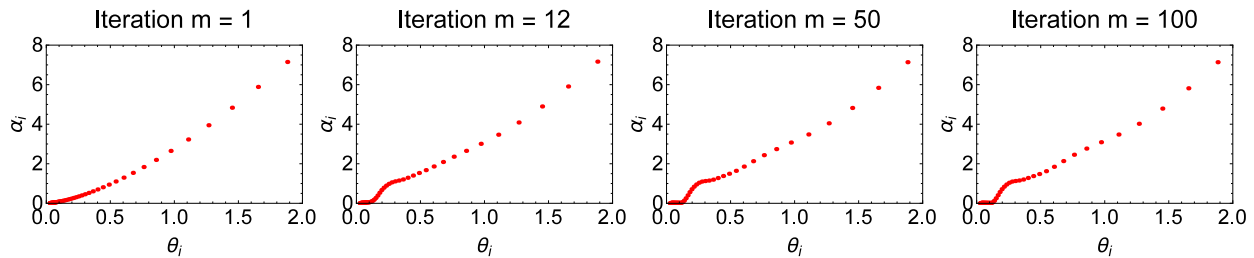}
\end{center}
\caption{Iterative matrix solution for the local shear viscosity $\alpha_i$ with the denoising function $\Psi$ versus function of the local reduced temperature $\theta_i$ at iteration number m = 0, 12, 50, 100. \label{fig:Iterations}}
\end{figure}
Fig.~\ref{fig:Iterations} shows the local shear viscosity $\alpha_i$ obtained using the denoising function,  versus the local reduced temperature $\theta_i$,  for different iteration numbers m.

\begin{figure}[H]
\begin{center}\
\includegraphics[width=7.0in]{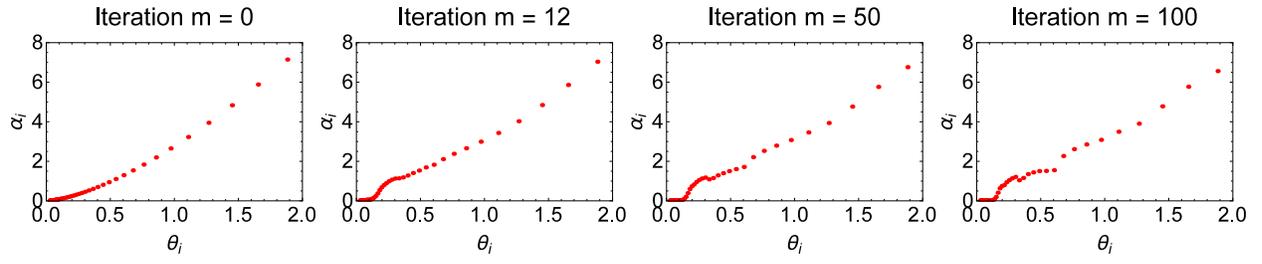}
\end{center}
\caption{Iterative matrix solution for the local shear viscosity $\alpha_i$ without the denoising function $\Psi$ versus function of the local reduced temperature $\theta_i$ at iteration number m = 0, 12, 50, 100. \label{fig:NoSmoothing}}
\end{figure}

Fig.~\ref{fig:NoSmoothing} shows $\alpha_i$ as a function of $\theta_i$ for different iteration numbers when the denoising function $\Psi$ is not utilized.    The denoising function $\Psi$ has very little influence on the iterative matrix solution at iteration number $m=12$, where the solution has converged.  At larger iteration numbers the stabilizing effect of $\Psi$ becomes apparent by comparing Fig.~\ref{fig:Iterations} and Fig.~\ref{fig:NoSmoothing}.  From this we can conclude that the $\Psi$ has a stabilizing effect on the algorithm at large iteration numbers, but does not significantly impact our determination of the local shear viscosity.

\end{document}